\pdfoutput=1

\def\singlecol{}
\let\singlecol\undefined 

\ifdefined\singlecol
  \documentclass[useAMS,usenatbib,referee]{aa}
\else
  \documentclass[useAMS,usenatbib]{aa}
\fi

\usepackage[pdftex]{}
\usepackage{times}
\usepackage{wasysym}
\usepackage{url}
\usepackage{lineno}
\usepackage{amssymb}
\usepackage{subfigure}
\usepackage{amsmath}

\usepackage{siunitx}

\usepackage{xcolor}

\usepackage{hyperref}
\hypersetup{
    colorlinks=true,
    linkcolor=magenta,
    citecolor=blue,
    filecolor=magenta,      
    urlcolor=cyan,
}

\newcommand{\etacar}{$\eta$~Car}
\newcommand{\g}{$\gamma$}

\newcommand{\fermi}{\emph{Fermi}-LAT}
\newcommand{\nustar}{\emph{NuSTAR}}
\newcommand{\hess}{\emph{H.E.S.S.}}

\newcommand{\cta}{\emph{CTA}}

\def\apjl{ApJ}%
\def\apjs{ApJS}%
\def\ao{Appl.~Opt.}%
\def\aap{A\&A}%
\DeclareSIUnit\solarmass{\ensuremath{M_\odot}}
\DeclareSIUnit\solarradius{\ensuremath{R_\odot}}
\DeclareSIUnit\year{yr}
\DeclareSIUnit\solarluminosity{\ensuremath{L_\odot}}
\DeclareSIUnit\gauss{G}
\DeclareSIUnit\erg{erg}
\DeclareSIUnit\pc{pc}

\def\@fnsymbol#1{\ensuremath{\ifcase#1\or *\or \dagger\or \ddagger\or
   \mathsection\or \mathparagraph\or \|\or **\or \dagger\dagger
   \or \ddagger\ddagger \else\@ctrerr\fi}}

\begin{document} 
\defcitealias{Ohm2015}{O15}

\title{Gamma-ray and X-ray constraints on non-thermal processes in $\eta$ Carinae}
\author{R.~White \inst{1,\star}  \and
        M.~Breuhaus \inst{1,\star} \and
        R. Konno \inst{2} \and
        S.~Ohm \inst{2} \and
        B. Reville \inst{1}
        \and J.A.~Hinton \inst{1} 
        }

\institute{Max-Planck-Institut f\"ur Kernphysik, Postfach 103980, D 69029 Heidelberg, Germany \\
           \and DESY, D-15738 Zeuthen, Germany}

\date{Accepted 2019 XX XX. Received 2019-10-31; in original form 2019-10-31}
  
\abstract
 {
  The binary system $\eta$ Carinae is a unique laboratory in which to study particle
  acceleration to high energies under a wide range of conditions,
  including extremely high densities around periastron. To date, no
  consensus has emerged as to the origin of the GeV \g-ray emission in
  this important system. With a re-analysis of the full \fermi\ dataset for $\eta$ Carinae we show that the spectrum is consistent with a pion decay origin. A single population leptonic model connecting the X-ray to \g-ray emission can be ruled out. 
  Here, we revisit the physical model of Ohm et al. (2015), based on two
  acceleration zones associated to the termination shocks in the winds of both stars.
  We conclude that inverse-Compton emission from in-situ accelerated electrons dominates the hard X-ray emission detected with \nustar\ at all phases away from periastron, and pion-decay from shock accelerated protons is the source of the \g-ray emission. 
  Very close to periastron there is a pronounced dip in the hard X-ray emission, concomitant with the repeated disappearance of the thermal X-ray emission, which we interpret as being due to the suppression of significant electron acceleration in the system.
Within our model, the residual emission seen by \nustar\ at this phase can be accounted for with secondary electrons produced in interactions of accelerated protons, in agreement with the variation in pion-decay \g-ray emission.
  Future observations with \hess, \cta\ and \nustar\ should confirm or refute this scenario.
  
}


\maketitle
\protect\footnotetext[1]{corresponding authors: Richard White, \email{richard.white@mpi-hd.mpg.de}, and Mischa Breuhaus, \email{mischa.breuhaus@mpi-hd.mpg.de}}
\section{Introduction} 
\label{sec:intro}

\etacar\ is the most prominent \g-ray emitting colliding wind binary system and is known to have many exceptional properties. The high mass-loss
rates of the binary stars in this system lead to very high densities in the wind collision region (WCR), which may in turn lead to efficient conversion of energy from
accelerated protons in to \g-rays and neutrinos.
The orbit of the system is highly eccentric with an eccentricity above \num{0.8} and most likely between \num{0.85} and \num{0.90}
\citep{Mehner_et_al2015} with a period of
$\sim\SI{5.5}{\year}$ \citep{Damineli2008}, which results in very different conditions in the WCR over the course of a single orbit. The primary star of the
system is believed to be a luminous blue variable \citep{Davidson1997}
and the companion a late-type nitrogen-rich O or Wolf-Rayet star
\citep{Iping2005}. Throughout the paper we will refer to the primary star as \etacar\ A and to the companion star as \etacar\ B. 
\etacar\ A has a mass larger than \SI{90}{\solarmass} with an initial mass of probably \SI{150}{\solarmass} or higher.
The mass of \etacar\ B is also not well known, but should be not larger than \SI{30}{\solarmass} \citep{Hillier2001}.
Their winds have different properties: the mass loss rate of \etacar\ A was estimated to be a few \SI{e-4}{\solarmass\per\year} up to a few \SI{e-3}{\solarmass\per\year} and the terminal wind velocity to be $\approx\SI{500}{\kilo\metre\per\second}$; for \etacar\ B possible values are around \SI{e-5}{\solarmass\per\year} and $\approx\SI{3000}{\kilo\metre\per\second}$ \cite[see][and references therein]{Corcoran_Hamaguchi2007}.
The surface magnetic-field strengths of the stars are uncertain. 
In discussing the radio measurements of WR147 made by \citet{Williams1997}, \citet{Walder2012} conclude that the emission is consistent with surface magnetic-field strengths of $30$ to $300$ G.\\

The variability of \etacar\ is well studied in the X-ray domain
\cite[see e.g.][]{Corcoran2005, Hamaguchi2007, Hamaguchi2014}.
An outline of the X-ray measurements between 1996 and 2014 is given in \citet{Corcoran2015} and put in perspective with the orbital dynamics of the system. The X-ray maximum is explained by a wind-wind cavity carved in the dense wind of \etacar\ A due to the motion of the companion $\eta$ Car B,  and its stellar wind. The X-ray minimum follows as a consequence of the subsequent blocking of the said cavity in further companion motion.\\

The first detections of \g-ray emission from the direction of \etacar\ were
obtained with the AGILE satellite \citep{Tavani2009} and \fermi\
\citep{FERMI:BSL,EtaCar:Fermi10}.
The \g-ray flux is observed to vary around the orbit, with an increase shortly before and a decrease around periastron \citep{EtaCar:Fermi10,EtaCar:Farnier11}, albeit less significantly than in the X-ray regime.
The \g-ray spectrum measured using \fermi\ can be described well
by two different components, one of which dominates the
emission above \SI{10}{\giga\electronvolt} \citep{EtaCar:Fermi10,EtaCar:Farnier11}.
The latter reference suggested that the low-energy component 
may be associated to inverse Compton (IC) scattering by electrons accelerated in the
WCR, and the second component due to decay of $\pi^0$s produced in the interactions of accelerated
hadrons in the dense environment.\\

\citet{Bednarek11} were the first to suggest that the two shocks in the system, owing to the different stellar wind properties for the individual stars, would lead to two populations of accelerated
particles with different maximum energies. They considered two different
models, one solely containing accelerated electrons, and the other both
high energy electrons and protons.
In \citet{Ohm2015} (hereafter \citetalias{Ohm2015}) this scenario was explored further with a more complete treatment
of the system geometry and dynamics. It was concluded that
the \g-ray emission is likely purely hadronic in origin,
with the two \fermi\ components indeed associated to the two shocks in the system.

The first full orbit of \etacar\ as
seen by \fermi\ was analyzed by \citet{Reitberger2015}. They found small variations at energies $< \SI{10}{\giga\electronvolt}$,
but larger ones for the high energy component, with a high flux at periastron, followed by a sudden decrease and a slow recovery. 
With the updated \fermi\ coverage of two periastron passages and
the major improvement in sensitivity (particularly at low energies)
represented by the {\it Pass 8} release of \fermi\ data, further information
about the variability in the \g-ray regime has been presented by \citet{Balbo2017}.
The high-energy component does not show a significant increase during the second
observed periastron passage, but the low energy component behaves in a
similar way during both passages. Comparing their results with \g-ray
fluxes derived from data of hydrodynamical simulations of
\citet{Parkin2011}, \citet{Balbo2017} argued again in favour of a model in which the low energy \g-ray emission is attributable to IC scattering.

Recently, \citet{Hamaguchi2018} have reported on observations with \nustar\ in the hard X-ray domain, which confirmed earlier suggestions \citep{Leyder2008}
of a hard, non-thermal, component at 30--50~keV.
These authors contend that this is IC emission connecting smoothly with the low energy component
seen with \fermi\,
inferring that the hard X-ray emission and the
low-energy \g-ray emission are produced by the same population of relativistic electrons.

Here we revisit the \fermi\ data, focusing on the previously unconstrained lowest energy ($<200$ MeV) part of the \g-ray spectrum with the aim of constraining the origin of this component (Section~\ref{sec:fermi}). In Section~\ref{sec:phys} we make an assessment of the likely origin of the different non-thermal components of \etacar\ based on 
physical considerations associated with particle acceleration, energy losses and the emission and absorption of
\g-rays.
In Section~\ref{sec:model} we present an update of the calculation of \citetalias{Ohm2015} and comparison to the newly available data.
Finally, in Section~\ref{sec:conc} we discuss the implications of the new data analysis and modelling, and the prospects for future observations.


\section{\fermi\ data analysis}
\label{sec:fermi}

\subsection{Data Selection}
\label{sec:fermi:data}

We have performed a \emph{Fermi}-LAT analysis following closely the approach taken by \citet{Reitberger2015} but considering: a wider energy range, additional data, an updated Galactic diffuse model, updated Instrument Response Functions (IRFs) and the latest \emph{Fermi}-LAT source catalogue. 

\emph{Fermi}-LAT data from the latest {\it Pass 8} release between August 4 2008 and July 1 2019 (MET 239557417 to 583643061) was analysed using {\it Fermitools} (version 1.0.7), the official Fermi ScienceTools data-analysis suite\footnote{The Fermi ScienceTools are distributed by
the Fermi Science Support Center (\href{http://fermi.gsfc.nasa.gov/ssc}{http://fermi.gsfc.nasa.gov/ssc}).} via {\it FermiPy} (version 0.17.4), an open-source python framework that provides a high-level interface to the Fermi ScienceTools; \citep{Fermipy2017}. 

Data were selected over a region of interest (ROI) of \SI{10}{\degree} by \SI{10}{\degree} centred at the nominal position of \etacar\ and aligned in galactic coordinates. Data were chosen according to the SOURCE event class ({\it evclass=128}) with FRONT+BACK event types ({\it evtype=3}). 

To minimise contamination from atmospheric $\gamma$-rays from the Earth, time periods in which the ROI was observed at zenith angle greater than \SI{90}{\degree} were excluded and the resulting exposure correction determined as part of the live-time computation, as per the current \fermi\ recommendations.\footnote{Note that in the past a cut on the rocking angle was also recommended, but is no longer deemed necessary.} A bright nova, ASASSN-18fv, approximately \SI{1}{\degree} away from \etacar\ was detected by the \emph{Fermi}-LAT in March 2018 \citep{asassn18}. To prevent contamination at the position of \etacar\ by emission from this nova, data in the period MET 542144904 to 550885992 were excluded. 

\subsection{Data Analysis}
\label{sec:fermi:an}

The \emph{Fermi}-LAT 8-year source catalogue (4FGL) \citep{4FGL2019} was used to construct a model of the sources surrounding \etacar. To account for the instrument point-spread-function and any extended sources just outside the chosen ROI, sources from an extended region of radius \SI{15}{\degree} were included in the model. In total \num{63} sources were included. In all cases the spectral model was included as-per the catalogue description. The latest Galactic diffuse background template and isotropic component files were utilised (see Appendix~\ref{sec:app-a} for the specific files used). \etacar\ was included  as it appears in the 4FGL catalogue under the identifier 4FGL~J1045.1-5940 as a point source with a log-parabolic spectral shape. 

A binned maximum-likelihood analysis using the recommended IRFs was performed in the energy range \SI{80}{\mega\electronvolt} to \SI{500}{\giga\electronvolt}. First the parameters of the background model were optimised in an iterative fashion, looping over all model components in the ROI fitting their normalisation and spectral shape parameters. Then a final fit was performed in which the galactic and isotropic background components remain fixed whilst the normalisation of all sources within \SI{3}{\degree} of \etacar\ with a Test Statistic (TS) $>\num{10}$ and at least \num{10} predicted events are allowed to vary. TS and residuals maps were then generated to assess the quality of the model. 

It was found that significant extended residual emission was present surrounding \etacar\ after this process. These residuals resemble the distribution of molecular gas in the region, suggesting an additional component associated with cosmic-ray interactions in the region. A template spanning \SI{1}{\degree} radius based on the CO survey of \citet{Dame2001} was generated and added as a diffuse component to the overall model, with a power-law spectral shape. The need for this additional diffuse component is discussed further in Appendix~\ref{sec:app-a}. The best overall model fit then results in a TS of $\sim\num{1000}$ for the new component, with a spectral index of $\num{2.1}\pm\num{0.1}$. The best-fit flux of this component is approximately \SI{60}{\percent} that of \etacar. The mean velocity-integrated intensity in the CO template corresponds to a total gas mass of approximately \SI{1e5}{\solarmass} assuming a conversion factor of CO to molecular hydrogen (X$_{\text{co}}$) of \SI{2e20}{\kelvin\kilo\metre\per\second} \cite[consistent with studies for this region, c.f.][]{Rebolledo2015}. The additional energy density in cosmic rays required to produce the energy flux of this component in \g-rays between \SI{100}{\mega\electronvolt} and \SI{30}{\giga\electronvolt} of \SI{7e-11}{\erg\per\square\centi\metre\per\second} is  $\approx\SI{1}{\electronvolt\per\cubic\centi\metre}$ over this $\sim\SI{80}{\pc}$ region. Whilst it is perhaps not surprising that the Galactic diffuse model used in \fermi\ cannot reproduce perfectly the emission from this complex region, this component is suggestive of an excess of locally accelerated cosmic rays in the region.
Although a model with multiple templates, allowing for different cosmic-ray densities in individual molecular clouds, may more accurately represent the diffuse emission, the approach taken here results in an acceptably flat residuals map, and is hence justified for use in the analysis of \etacar. The resulting model including this component was used as a starting point for the further analysis steps outlined below.

\subsection{Spectral Energy Distribution}
\label{sec:fermi:sed}

A spectral analysis of the full \fermi\ data set was performed based on the optimised model described above. The resulting SED (black points) together with systematic-error band (grey area) are shown in Figure~\ref{fig:sed}. For details of the systematic error estimate see Appendix~\ref{sec:app-a:sys}. The SED for the total \etacar\ data set from this work can be seen to be consistent with that from \citet{Reitberger2015} (light blue). A drop in flux below \SI{300}{\mega\electronvolt} is clearly visible in the measured SED, indicative of a kinematic pion-cutoff. The naima package~\citep{Naima} was used to fit the SED points under the assumption that the \g-ray emission arises entirely from proton-proton interactions (black curve). The measured SED appears inconsistent with a purely leptonic scenario, such as that presented in \citet{Hamaguchi2018}. 

\begin{figure}
  \centering 
  \ifdefined\singlecol
    \includegraphics[width=0.8\textwidth,draft=false]{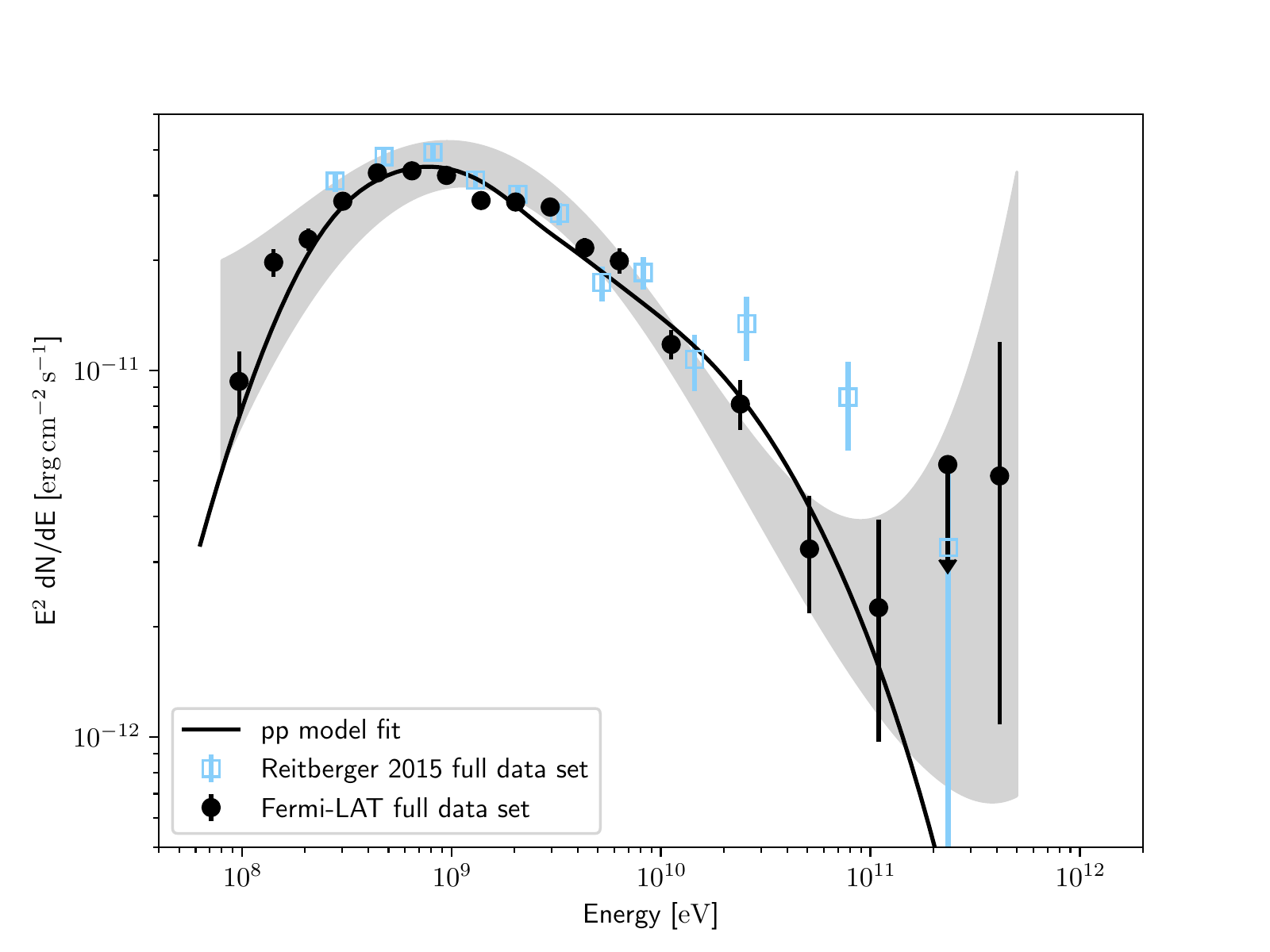} 
  \else
    \includegraphics[width=0.495\textwidth,draft=false]{./plots/figure1_v2.pdf}
  \fi
  \caption{The SED resulting from \fermi\ analysis of \etacar\ over the full data set (black points) together with a systematic-error band (grey area). A fit to the data points below \SI{200}{\giga\electronvolt} using the Naima package~\citep{Naima} is shown by the black curve under the assumption that the \g-ray emission arises from proton-proton interactions with an exponential cutoff powerlaw distribution for the protons. For comparison data points from a previous \fermi\ analysis \citep{Reitberger2015} are shown in blue.}
  \label{fig:sed}
\end{figure}

\subsection{Phase-Resolved SED Analysis}
\label{sec:fermi:phasesed}

The \emph{Fermi}-LAT data on \etacar\ spans \num{11}~years covering two periastron passages over a relative phase range of \num{0.92} to phase \num{2.88}. SEDs were produced for the following phase ranges motivated by the X-ray behaviour:
\begin{itemize}
\item{\bf Pre-Periastron:} Phase ranges \num{0.92} - \num{0.99} and \num{1.92} - \num{0.99} (approximately \num{140} days each) covering the gradual build up in flux observed in the X-ray light curve. 
\item{\bf Periastron:} Phase ranges \num{0.995} - \num{1.025} (approximately \num{60} days each) covering the abrupt dip in flux observed in the X-ray light curve. 
\item{\bf Off-Periastron:} Phase ranges \num{1.1} - \num{1.9} and \num{2.1} - \num{2.88}, excluding \num{2.65} - \num{2.70} for the nova, ASASSN-18fv \citep{asassn18} where no significant variability is observed in the X-ray light curve.
\end{itemize}

In each phase range a fit was performed starting from the optimised model in which the Galactic and isotropic background components remain fixed to the previous optimised values. The normalisation of all sources within \SI{3}{\degree} of \etacar\ with a TS $>\num{10}$ and at least \num{10} predicted events were allowed to vary. The analysis was performed both independently for each orbital period and with both orbital periods combined. The SEDs from the combined analysis of both orbits are presented in Section~\ref{sec:phys} and compared to that from the individual orbital periods in Appendix~\ref{sec:app-a:sed}.

\subsection{Temporal Analysis}
\label{sec:fermi:lc}

\begin{figure}
  \centering 
  \ifdefined\singlecol
    \includegraphics[width=0.8\textwidth,draft=false]{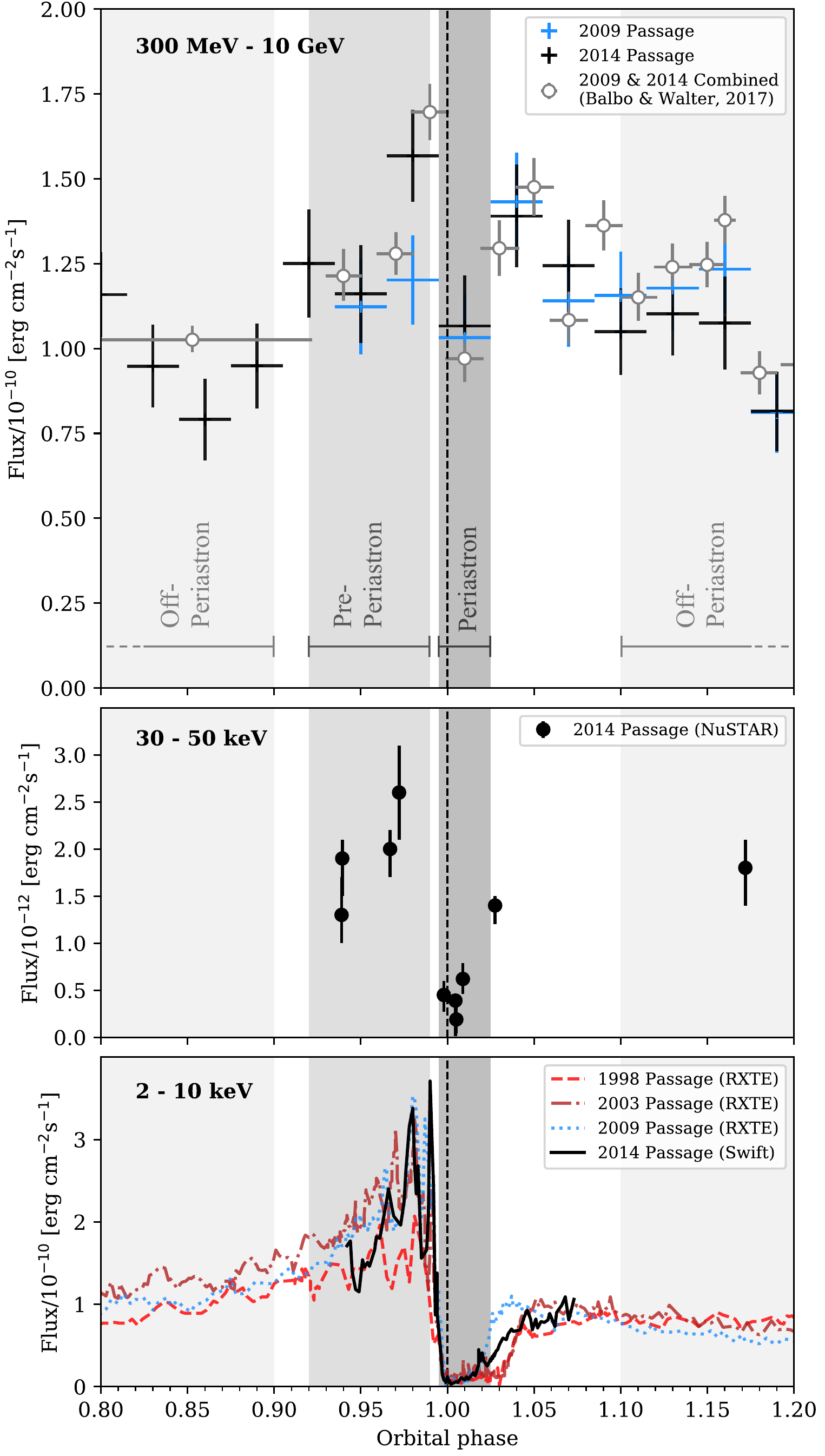} 
  \else
    \includegraphics[width=0.495\textwidth,draft=false]{./plots/fermi_lc.pdf}
  \fi
  \caption{{\bf Upper panel:} \emph{Fermi}-LAT temporal analysis of \etacar\ from \SI{300}{\mega\electronvolt} to \SI{10}{\giga\electronvolt} from orbital phase \num{0.8} to \num{1.2} in \num{0.03} phase bins. Data corresponding to the first orbital period observed by the \emph{Fermi}-LAT (with periastron in \num{2009}) are shown in blue, and for the second (with periastron in \num{2014}). Vertical grey shaded areas indicate the phase ranges used in the spectral analysis presented in Section~\ref{sec:fermi:phasesed} around periastron, pre-periastron and off-periastron. The points from \cite{Balbo2017} are shown as hollow grey circles. The vertical black dashed line indicates periastron. {\bf Middle panel:} The \nustar points for the \num{2014} periastron passage from \citep{Hamaguchi2018}. {\bf Lower panel:} RXTE and Swift data for the last four periastron passages. 
}
  \label{fig:lc}
\end{figure}

The upper panel of Figure~\ref{fig:lc} shows the light curve around periastron for the \fermi\ data in the energy range \SI{300}{\mega\electronvolt} to \SI{10}{\giga\electronvolt} divided into \num{0.03} bins in orbital phase (approximately \num{60} days) and aligned to the periastron phase range used in the SED analysis described in the previous Section. The light curve was produced by performing an independent likelihood analysis in each time bin starting from the optimised model described in Section~\ref{sec:fermi:an}. In each time bin only nearby source normalisations and the normalisation of \etacar\ were allowed to vary. An energy range of \SI{300}{\mega\electronvolt} to \SI{10}{\giga\electronvolt} in-which the spectral shape of \etacar\ is roughly constant was chosen. Data from the first orbital period observed by \emph{Fermi}-LAT (with periastron in \num{2009}) are shown in blue whilst data from the second orbital (periastron \num{2015}) are shown in black. For comparison the points from \cite{Balbo2017} are included. The hard X-ray emission as measured by NuSTAR during the \num{2014} periastron passage \citep{Hamaguchi2018} is shown in the middle panel. In the lower panel the X-ray emission at lower energies as measured by RXTE and Swift for passages culminating with periastron in \num{1998}, \num{2003}, \num{2009} and \num{2014} are shown \citep{Corcoran2017}. The \SI{300}{\mega\electronvolt} to \SI{10}{\giga\electronvolt} \g-ray flux behaves similarly during both passage (as noted by \cite{Balbo2017}) and shows no distinct disappearance at periastron as seen in the X-ray data.

\section{Physical considerations and limits}
\label{sec:phys}
Before describing the numerical model, many important conclusions can be drawn about the significant effects and likely origin of different radiation components in the \etacar\ system. From estimates of stellar luminosities, mass-loss rates, wind velocities and surface magnetic fields, order of magnitude estimates of the conditions at the two shocks within the system can be derived and associated acceleration and energy-loss timescales considered. In this section we consider in turn each of the primary physical processes at work in the system, which have implications for the non-thermal emission.\\

Following \citet{Bednarek11}, we consider the WCR separated from the winds of the two stars on either side by shock fronts. Despite the large densities, both shocks are collisionless 
\cite[Hall parameter $\omega_{\rm gi} \tau_{\rm ii}$ of the order $10^5$ or larger, c.f.][]{Eichler1993}, 
and in principle provide sites for the acceleration of energetic particles via diffusive shock acceleration. The conditions pertinent to this process are discussed below.

{\bf Acceleration at \etacar\ A's shock:} The inferred high surface fields and terminal velocity of the wind of $\eta$ Car A limits the shock's Alfv\'en Mach number to rather modest values, $M_{\rm A} < \num{10}$, reaching a minimum near periastron. At all phases, the inferred low Mach numbers restrict the amount of self-generated MHD fluctuations that in-situ accelerated particles can amplify upstream of the shock. From energy conservation, we must satisfy ${\delta B^2}/{4 \pi} = \chi P_{\rm cr}$, where $P_{\rm cr}$ is the cosmic-ray pressure, and $\chi\ll 1$ the fraction that is converted to magnetic pressure. Re-arranging as 
\begin{equation}
    \frac{\delta B^2}{B_0^2} = \chi \left(\frac{P_{\rm cr}}{\rho v_{\rm sh}^2}\right) M_{\rm A}^2 \, ,
    \label{Eq:Bmax}
\end{equation} 
and using the results of \citet{Bell04}, where $\chi \approx v_{\rm sh}/c$ we find only modest growth of magnetic fluctuations ($\delta B \lesssim B_0$) is possible.
In the most optimistic scenario, the Hillas limit \citep{Hillas} for the shock of \etacar\ A is (neglecting losses)
\begin{equation}
E_{\rm Hillas} = 50 \left(\frac{B_{\star}}{100{\rm G}}\right)
\left(\frac{R_{\star}}{100 R_\odot}\right)
\left(\frac{\theta_A}{60^\circ}\right)
\left(\frac{v_{\rm rot}}{0.1 v_\infty}\right)~
{\rm TeV}  \enspace, \nonumber 
\end{equation}
where $R_{\star}$ and $B_{\star}$ are the stellar radius and surface magnetic field respectively, $L \approx r_{\rm sh} \theta_A$ is the spatial extent of the shock, and for convenience we have taken a purely toroidal magnetic field in the wind, $B\propto r^{-1}$. However, as we will show below, the maximum energy in \etacar\ A is always limited by losses. 
We note that while significant magnetic-field amplification can be ruled out, the growth rate for the classical ion-cyclotron instability \cite[e.g.][]{Tademaru} can be sufficiently rapid to ensure acceleration efficiency close to the Bohm limit. 

{\bf Acceleration at \etacar\ B's shock:} The faster wind of $\eta$ Car B should result in large Alfv\'en Mach numbers.  Making the reasonable assumption that the star rotates at $ <\SI{10}{\percent}$ of its critical velocity, the shock will be quasi-parallel ($<\SI{45}{\degree}$) over its entire orbit. The Alfv\'en Mach number will thus increase monotonically with distance from the star. Plausible stellar surface parameters and wind conditions imply Alfv\'en Mach numbers  $M_{\rm A} > \num{100}$, and consequently the possibility of non-linear magnetic-field amplification should be considered, c.f. Equation~(\ref{Eq:Bmax}). 

An estimate for the maximum attainable energy can be found from the Hillas limit, assuming strong field amplification. In reality, the statistics of the fields plays an important role, and the Hillas result should be considered here as a firm upper limit. Adopting standard parameters for the saturated magnetic field \cite[e.g.][]{Bell04}:
$P_{\rm cr}/\rho v_{\rm sh}^2 =0.1$, $\chi =v_{B,\infty}/c$; the Hillas limit gives $E_{\rm Hillas} \approx\SI{30}{\tera\electronvolt}$. Hence, $\eta$ Car can be seen to provide a unique time-dependent laboratory for testing theories of cosmic-ray acceleration with self-amplified fields relevant to high Alfv\'en Mach number supernova remnant shocks for example. 
Conversely, as the resulting \g-ray emission is dependent on the physical characteristics of the stellar wind, the high-energy observations may provide new information on what is otherwise a poorly constrained star.

{\bf Electron heating and injection:} The shock conditions close to periastron require special consideration, as the injection of electrons is sensitive to shock obliquity and Mach number. In the case of $\eta$ Car A, 
the relatively modest Alfv\'en Mach number may restrict electron injection, unless the shock is highly oblique. Hence, electron acceleration may not occur close to the star, where the fields are predominantly radial. Close to periastron, the shock may even become sub-critical \citep{EdmistonKennel}, which can suppress also the electron heating. 

At $\eta$ Car B, again the situation is different, although electron acceleration may also be inhibited close to periastron. The fate of the shock close to periastron is in fact unclear. For example, it has been argued that the shock of $\eta$ Car B will collapse onto its stellar surface during closest approach \cite[e.g.][]{Parkin2011}, although this requires a very weak surface magnetic field. As we argue below, the sustained \g-ray emission requires persistence of the shock. Another factor that can affect the non-thermal electron spectrum concerns the fact that for quasi-parallel shocks, electron acceleration requires the shock velocity to exceed the whistler phase speed \citep{Eichler1993}, i.e. $M_{\rm A} \gg \sqrt{m_i/m_e}$.
Since $M_{\rm A} \propto r$ in the radial field zone, this condition will be violated close to the star unless the surface fields are extremely weak. 

{\bf Interactions of protons:} Accelerated protons and nuclei inevitably interact with the thermal background particles, leading to the production of \g-rays, neutrinos and secondary leptons, which may in turn produce X-rays and \g-rays via the IC process. For \etacar\ A, the proton energy-loss timescale via such collisions at the stagnation point is of the order of \SI{10} days,
significantly shorter than the propagation timescales via advection or diffusion. This is therefore a calorimetric system, in which energy gained by acceleration is lost in-situ and the maximum energy attainable can be 
estimated from equating the timescales for energy loss $(t_{\rm pp}\approx \left(n_{\rm gas}\sigma_{\rm pp} c)^{-1}\right)$ and diffusive shock acceleration $(t_{\rm acc}\approx \eta_{\rm acc} r_{\rm g} c / v_{\rm sh}^2)$: 
\begin{eqnarray}
    E_{\text{max}} \approx
    70 ~\eta_{\rm acc}^{-1}\left(\frac{v_{\rm sh}}{10^3~{\rm km~s}^{-1}}\right)^2\left(\frac{B}{1~{\rm G}}\right)
    \left(\frac{r}{{\rm AU}}\right)^2 ~{\rm GeV}\enspace .
\end{eqnarray}

For our assumed orbital parameters, and a constant acceleration efficiency of $\eta_{\rm acc}=\num{5}$, this gives mean values of $\sim$ \SI{150}{\giga\electronvolt}, but the energies are always below \SI{230}{\giga\electronvolt}.
At the shock in the wind of \etacar\ B, densities are much lower and velocities higher. The flow timescales are shorter than
the energy-loss timescales and it seems plausible that the maximum energy is determined by the residence time of protons and nuclei in the acceleration region, potentially reaching its Hillas limit
\cite[e.g.][]{AchterbergHouches}. The fraction of these particles that interact rather than escape from the system is dependent on the details of the cosmic-ray transport out of the system and also on the way in which the material in the two very different stellar winds becomes mixed on larger scales \citepalias[see also][]{Ohm2015}. 

{\bf Electron cooling:} In \etacar, conditions are such that energy-loss timescales for relativistic electrons are always much shorter than orbital timescales or particle escape timescales. An equilibrium between particle acceleration and cooling is therefore rapidly established at all electron energies.
The dominant cooling process at the lowest energies is through electronic excitations \citep{Gould75}, at higher energies IC losses dominate and finally for the highest-energy electrons Klein-Nishina suppression beyond $\sim\SI{10}{\giga\electronvolt}$ energies leave synchrotron emission as the dominant cooling process \citepalias[see Figure 2 in][]{Ohm2015}.
The transition point between Coulomb-Ionisation losses and IC cooling depends on the ratio of thermal electron density to radiation energy density. As both of these quantities vary as the inverse square of the distance to the star, this cross-over point is approximately constant with phase. Following \citet{Schlickeiser2002} the transition occurs at electron energies of:

\begin{equation}
\begin{split}
    E_{\text{Break}} \approx   \num{130}   \left(\frac{\dot M}
    {\dot M_{\text{A}}}\right)^{\frac 1 2}  
     \left(\frac{v_{\infty}}{v_{\infty,\text{A}}}\right)^{-\frac 1 2}\left(
    \frac{L}{L_{ \text{A}}}\right)^{-\frac 1 2} \si{\mega\electronvolt},
\end{split}
\label{eq:break}
\end{equation}
i.e. the break energy is \SI{130}{\mega\electronvolt} for the side of \etacar\ A and \SI{30}{\mega\electronvolt} for \etacar\ B. The subscript \lq A\rq\ in Equation~(\ref{eq:break}) refers to the properties of \etacar\ A.
More accurate calculations including the radiation fields from both stars and averaging over phases reduces these values to
\SI{110}{\mega\electronvolt} and \SI{15}{\mega\electronvolt}
respectively. In equilibrium the electron spectrum will in any case soften above this point, leading to a break in the resulting IC spectrum, 
which lies well below the \fermi\ energy range for any reasonable choice of parameters. 

{\bf Anisotropy of IC emission:} As the typically assumed situation of isotropy in both the electron and target photon populations is not met in \etacar, it is important to evaluate the impact of anisotropy on the expected IC emission. The electrons can still be assumed to be isotropic, since the velocities of the non-thermal particles are much larger than the flow velocities.
We therefore adopt the cross section of \citet{Aharonian1981} for the anisotropic IC scattering of an isotropic electron distribution. We find that there is a general compensating effect when one accounts for the radiation fields of the two stars and the different locations in the shock cap, such that the impact of anisotropy on the flux is always less than \SI{30}{\percent} for the energy range \SIrange{30}{50}{\kilo\electronvolt} where IC emission from electrons dominates and measurements from \nustar\ are available. Averaging over different phases reduces the impact of anisotropy even further. 

{\bf \g-\g\ pair-production:} Above a threshold of $\sim\SI{30}{\giga\electronvolt}$, \g-rays inside the \etacar\ system can pair-produce in the stellar radiation fields. The angular-dependent effects lead to phase-dependent flux variability in the VHE domain. In the core energy of \fermi\ around \SI{1}{\giga\electronvolt} \g-\g\ absorption is negligible. In our detailed model, described in the next section, we consider \g-\g\ pair-production only as an absorption process. Here however we note that the presence of a significant flux of \g-rays beyond \SI{30}{\giga\electronvolt} inside the system may result in a measurable contribution to the IC emission from the resulting pairs \citep[see e.g.][]{Dubus2006,Bosh-Ramon_Khangulyan2009}.
The IC emission of the pair-produced particles is expected to peak between \num{5} and \SI{10}{\giga\electronvolt} and results in an enhancement of $\sim\SI{10}{\percent}$ close to periastron, due to the proximity of the stars and the smaller size of the emission region. At all other phases the enhancement is less than \SI{1}{\percent}.

\begin{figure*}[!h]
    \centering
    \includegraphics[width=0.8\textwidth]{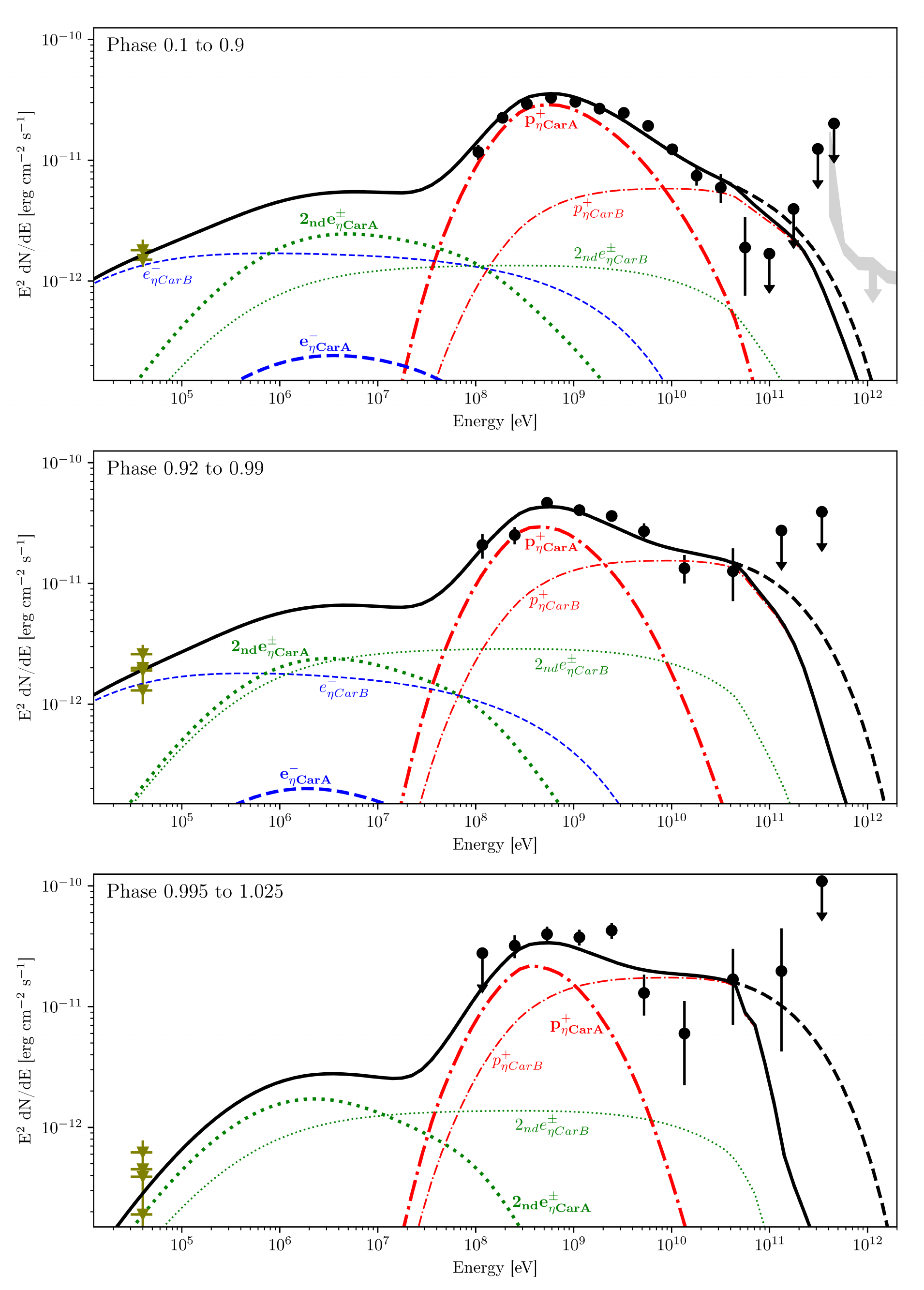}
    \caption{SEDs for the phase ranges off-periastron (top), pre-periastron (middle) and periastron (bottom). Black data-points represent the points from our \fermi\ analysis (Section~\ref{sec:fermi:phasesed}), olive points show data from \nustar\ \citep{Hamaguchi2018}. The black solid line shows the total emission from our model with absorption. The black dashed line shows the same without absorption. Blue lines show the emission from electrons, green lines the contribution from secondary particles and red lines from protons of \etacar\ A (bold) and \etacar\ B (thin). In the upper plot (phase 0.1 to 0.9) upper limits from \hess\ were added in grey \citep{Abramowski2012}.}
    \label{fig:modelSEDs}
\end{figure*}


\section{Phase-dependent Model}
\label{sec:model}

For the modelling of the non-thermal emission of \etacar\ as a function of phase we make use of the framework presented in \citetalias{Ohm2015}, with some refinements and adjustment of parameters for consistency with the new full \fermi\ data-set as described above. 

\subsection{Modelling Approach}

We model diffusive shock acceleration at the two shocks in the system, formed on either side of the contact discontinuity between the two stellar winds \citep{Eichler1993}. We assume constant and spherically symmetric winds, with the geometry of the shock cap established by the momentum balance between the two winds (which are assumed to reach terminal velocity before the shock at all phases), see \citet{Canto1996}. The parameters of the two stars and their winds are given in Table~\ref{tab:pars}. Acceleration is assumed to take place with Bohm scaling (mean free path proportional to gyroradius), using the scaling factor $\eta_{\rm acc}$, which is a free parameter of the model.

We mimic the effect of magnetic-field amplification at the shock in the wind of \etacar\ B by adopting a large rotation velocity. That this is reasonable follows from Equation~(\ref{Eq:Bmax}), where one can see that for constant acceleration efficiency, the amplified field corresponds to $\delta B^2 \propto \rho$. Provided anisotropic transport is unimportant, which is implicitly assumed in the model, the magnetic-field amplification is equivalent to a wind with a purely toroidal field. 

Electron acceleration in the system is found to be limited by IC losses at all phases and everywhere in the system. The loss timescales at all energies are also short compared to propagation timescales and so the local equilibrium electron spectrum can be calculated based on the local environment in each of the considered cells (radial and azimuthal bins on the shock surface, see \citetalias{Ohm2015}) at each of the two shocks. 

For the high-density/low-velocity shock associated with \etacar\ A, proton acceleration is found to be limited at all phases by collisions with ambient nuclei. The \g-ray spectrum and injection spectrum of secondary electrons in each cell of the shock cap is calculated assuming an equilibrium between acceleration and these losses. The equilibrium IC spectrum of the secondaries is then found in each cell, again assuming in-situ cooling.

For the lower-density/higher-velocity shock associated to \etacar\ B the situation is more complex. Acceleration of protons is limited by the residence time of particles in the system, which is assumed to be dominated by advection downstream of the shock. The changing conditions during acceleration, and an approximation of the effect of CR modification to the shock are dealt with as in \citetalias{Ohm2015}. After leaving the acceleration region particles are assumed to travel ballistically out of the system. This point is defined as the ballistic point. Hadronic interactions may take place in the shock cap, but primarily occur in a zone in which mixing between the shock-compressed winds from the two stars is assumed to take place. The scale on which this mixing takes place is a free parameter of the model, with the adopted value of one shock cap radius. The shock cap radius is defined as the distance between the apex and the ballistic point. The value for the mixing length is motivated by hydrodynamical models \cite[e.g.][]{Parkin2011}.

For this work we have adopted a more complete treatment of secondary particles, in particular secondary production in the ballistic outflow beyond the shock cap which was neglected in \citetalias{Ohm2015}. Secondaries are again assumed to cool in-situ, with the local equilibrium spectrum calculated based on the environment in each bin in distance along the trajectory of each azimuthal bin, during each phase step. 

The effect of \g-\g\ interactions in the system is modelled as in \citetalias{Ohm2015}, with the simplifying assumptions of point-like emission from the two stars and neglecting the radiation of the $e^{+}/e^{-}$ pairs produced.

A major change here is the assumption concerning the shock cap behaviour during the periastron passage. In our previous work we considered the disappearance of part of the shock cap and both of the associated shocks during the period in which thermal X-ray emission is absent. 
Here, motivated by the considerations given in Section \ref{sec:phys} above, we assume instead that both shocks persist but that the injection/heating of electrons is suppressed at phase \num{0.995} - \num{1.025}, resulting in both the disappearance of the thermal X-ray emission and a halt to the injection of electrons in to the shock acceleration process.
We note that the shocks can not vanish completely as the proton loss times are shorter than the phase range in question, implying continuous acceleration of nuclei is essential. 

The power used for acceleration of CRs depends on the local kinematics. In each cell with azimuthal and radial bin $(i,j)$, the derived CR power $P_{\text{cr},(i,j)} = \epsilon P_{\text{wind},(i,j)}$ is given as a fraction $\epsilon$ of the locally available kinetic wind power
 
 \begin{eqnarray}
    P_{\text{wind},(i,j)} =
    \frac{\dot{M}v_\infty^2}{2} \left(\frac{v_{n_{(i,j)}}}{v_\infty}\right)^2
   \frac{\mathit{\Delta\Omega_{(i,j)}}}{4\pi}\enspace .
\end{eqnarray}
 
 Here, $v_{n_{(i,j)}}$ is the normal velocity component to the shock front and $\mathit{\Delta\Omega_{(i,j)}}$ is the solid angle fragment of the local injection area on the shock cap. The efficiency $\epsilon$ is a free parameter of the model. In equilibrium, where particles are accelerated up to their cooling limit in a single cell,
 the CR power is determined at injection. In the system of \etacar\ B, however, the protons do not reach equilibrium and accelerate during propagation through the shock cap  \citepalias[see][]{Ohm2015}. Here, the CR power will increase until the particles leave the accelerating region.

\subsection{Modelling Result}
The key results of the modelling are summarised in 
Figure~\ref{fig:modelSEDs}, where our 
model is compared to the new \fermi\ analysis and existing \nustar\ data. 
The phase ranges \numrange{0.1}{0.9} (off-periastron, upper panel), \numrange{0.92}{0.99} (pre-periastron, middle panel) and \numrange{0.995}{1.025} (periastron, lower panel) are shown. The model parameters used 
are provided in Table~\ref{tab:pars}. 
\nustar\ datapoints in the corresponding phase ranges from \cite{Hamaguchi2018} are shown in olive. Black points show the results from our analysis of the \fermi\ data (see Section \ref{sec:fermi:phasesed}). We also provide in the top panel, the far-from-periastron upper limits from \hess\ \citep{Abramowski2012}.  The parameters required to explain the \fermi\ emission as arising from pion-decay are very similar to those of \citetalias{Ohm2015}. We found that values of $\eta_{\text{acc}}=\num{15}$ for \etacar\ A and $\eta_{\text{acc}}= \num{5}$ for \etacar\ B, while putting $\approx\SI{10}{\percent}$ of the available wind power at the shock cap into the acceleration of protons above a \si{\giga\electronvolt} matches the \fermi\ data well for all phase ranges considered. 
The solid black line shows the total emission from the model and the dashed black line at higher energies shows the total emission without absorption. Different emission components are shown in different colours: red for the emission from protons, blue for primary electrons and green for emission from secondary particles created in proton-proton collisions.

At off-periastron we find that primary electrons (indicated with blue lines in Figure~\ref{fig:sed}) are required to explain the level of emission seen with \nustar. We fix the relative fraction of energy going into primary electrons above a \si{\mega\electronvolt} to that in protons above a \si{\giga\electronvolt} at all phases for the two stars to be \SI{3}{\percent}. IC emission from electrons accelerated at the side of \etacar\ B dominates at \nustar\ energies, even if the fraction of energy going into the electrons of \etacar\ A is increased by one order of magnitude.
Pion decay from proton interaction on the side of \etacar\ A is, in all phase ranges, the main contributor to the low energy \fermi\ component of the SED, whereas at higher energy, protons from \etacar\ B can account for the observed flux and spectrum.
At periastron we switched off the injection of electrons. The level of the \nustar\ emission is consistent with the emission from secondary particles and no primary electrons are required in our model to match the data. This might be a hint for the consideration mentioned in Section \ref{sec:phys} that at periastron the nature of the shock is changed, such that electron acceleration is quenched.

The  spectral breaks at lower energies in the IC emission are associated to the transition between Coulomb and IC cooling (see Equation~\ref{eq:break}) and occur at different energies due to the large range of different densities associated to the two shocks and the mixing region beyond the shock cap.

Absorption is important for energies above \SI{100}{\giga\electronvolt} and therefore affects mainly the emission from \etacar\ B. Stronger absorption occurs during periastron compared to the other phases due to the small spatial extent of the emission region and the proximity to the two stars. However, this would change with the size of the emission region which depends on the mixing length scale in the ballistic region.

The emission from protons of \etacar\ B increases more towards periastron than for \etacar\ A. The reason for this is the escape of accelerated protons at the side of \etacar\ B, which do not contribute to the \g-ray emission. Towards periastron the density increases and more particles can interact.

\begin{table}
    \centering
    \begin{tabular}{l c c}
    \hline \\
    Parameter   & \etacar\ A    & \etacar\ B   \\ \\
    \hline \\
    $R_{\star}$ [\si{\solarradius}]     & \num{100}   & \num{20}     \\
    $T_{\star}$ [\si{\kelvin}]    &  \num{2.58e4}   & \num{3e4}       \\
    $L_{\star}$ [\SI{e6}{\solarluminosity}] & \num{4}   & \num{0.3} \\
    $\dot M$ [\si{\solarmass\per\year}] & \num{4.8e-4}  & \num{1.4e-5}\\
    $v_{\infty}$ [\si{\kilo\meter\per\second}]  & \num{5e2} & \num{3e3} \\  
    $B_{\star}$ [\si{\gauss}]  & \num{100} & \num{100} \\
    $v_{\text{rot}}$ [$v_{\infty}$] & \num{0.15}    & \num{0.15} \\
    $\eta_{\rm acc}$          & \num{15}  &   \num{5} \\
    $P_\text{p}$ [$P_\text{wind}$]    & \SI{10}{\percent}  & \SI{9}{\percent}\\
    $P_\text{e}/P_\text{p}$ &  \SI{3}{\percent}   & \SI{3}{\percent}   \\
    \\ \hline
    \end{tabular}
    \caption{Adopted model parameters associated with the two stars in the \etacar\ system and their associated wind shocks. For $T_{\star}$, $L_{\star}$, $\dot M$ and $v_{\infty}$ we used the same values as listed in Table 4, and for
     $R_{\star}$ the same values as in Table 1  of \cite{EtaCar:Parkin09}. The surface magnetic fields $B_{\star}$, the rotation velocities $v_{\text{rot}}$, the acceleration efficiency parameters $\eta_{\rm acc}$, the wind power going into the acceleration of protons  $P_\text{p}$ and the ratio $P_\text{e}/P_\text{p}$ of the power going into the acceleration of electrons above \SI{1}{\mega\electronvolt} and protons above \SI{1}{\giga\electronvolt} were parameters adopted for our model (see text for details).}
    \label{tab:pars}
\end{table}


\section{Discussion and Conclusions}
\label{sec:conc}

Based on the new \fermi\ analysis, the basic physical arguments of Section~\ref{sec:phys} and the SED modelling of Section~\ref{sec:model}, several important conclusions on the non-thermal processes in \etacar\ can be reached. Here we consider each of these in turn.

{\bf Pion-decay origin of the GeV emission:} The spectral shape of the emission seen with \fermi\ between \num{0.1} and \SI{10}{\giga\electronvolt}, as shown in Figure~\ref{fig:sed}, strongly favours a pion decay origin. This conclusion is consistent with our previous arguments in \citetalias{Ohm2015}, based on simple energetic considerations of a two-shock system. 
In our model, the emission at the lower end of the \fermi\ band is dominated by protons accelerated at the shock of \etacar\ A, see Figure~\ref{fig:modelSEDs}. The maximum energy at the shock is limited by hadronic interactions in the dense wind. The lower densities associated with the shock on the side of \etacar\ B allow for larger maximum energies, the latter being limited only by escape from the shock cap. These protons thus account for the high energy tail of the \fermi\ measurements, where the \g-rays are produced primarily as the shocked plasma from \etacar\ B, with entrained high-energy particles, mixes with the denser material from \etacar\ A in the exhausts of the WCR. 
The detailed physics of the mixing will affect not only the cooling of protons and resulting \g-ray emission, but also the cooling break of secondary pairs. We have found that the simple mixing-length profile adopted in our model provides a reasonable match to the data, but a more sophisticated model may be necessary to capture specific features.\\

{\bf Leptonic emission:} 
Whilst there is a consensus that emission above \SI{10}{\giga\electronvolt} is produced in hadronic interactions, the origin of the emission between \SIrange{0.1}{10}{\giga\electronvolt} was still under debate \cite[e.g. \citetalias{Ohm2015};][]{Reitberger2015,Balbo2017,Hamaguchi2018}.
The latter authors have argued that the emission observed at these energies might represent a smooth continuation of the hard spectrum emission seen at 10s of \si{\kilo\electronvolt} arising from the IC emission of a hard electron spectrum. This scenario is strongly disfavoured by the new low-energy \fermi\ data points. In principle, as the IC emission from \etacar\ B's primary electrons can account for the \nustar\ data, a significantly harder lepton population accelerated at \etacar\ A could be contrived to match the low-energy end \fermi\ data. However, this would require an equilibrium spectrum \emph{harder} than $E^2 dN/dE \propto E^{1/2}$, which is incompatible with standard DSA theory.
It is also difficult to disregard the coincidence between the position of the low-energy roll-over at $\lesssim\SI{300}{\mega\electronvolt}$ with the threshold energy for pion production.

Finally, the strong orbital variability observed with \nustar\ \citep{Hamaguchi2018}, in contrast with the modest variability in the \g-ray flux, is also difficult to reconcile with a single electron population producing the emission in both energy bands. 

{\bf \nustar\ observations:} The non-thermal component in the \nustar\ data \citep{Hamaguchi2018} confirms the presence of primary electrons.
As discussed in Section~\ref{sec:phys}, the equilibrium electron spectrum in the system must include a break associated to the transition from dominant Coulomb/Ionisation to IC cooling. For \etacar\ A, the break in the IC spectrum is at a few \si{\mega\electronvolt}, while for \etacar\ B it can fall close to the \nustar\ band (see for example \citet{HHARAA} or directly from the IC breaks in Figure~\ref{fig:modelSEDs}). In any case, the IC spectrum is expected to soften between the \nustar\ and \fermi\ bands, which is consistent with the hard \nustar\ spectrum. 

    Far from periastron, the \nustar\ flux is compatible with IC emission originating almost exclusively from the primary electrons accelerated at the shock of \etacar\ B. In this regard, the fractional power injected into electrons at \etacar\ A is effectively unconstrained. 
    The \nustar\ minimum around periastron is in agreement with the level of emission expected for secondary electrons. The reason for the absence of primary electrons during this phase is not yet certain and will require more detailed kinetic simulations. However, the persistence of the \fermi\ \g-ray flux (see Figure~\ref{fig:lc}) is a clear indication that both shocks at the edges of the WCR are continuously accelerating protons. The absence of measurable X-ray flux from primary electrons therefore is most probably a consequence of suppression of electron heating/scattering, which we interpret as being due to low Alfv\'en Mach number in the case of \etacar\ A, and a runaway of the whistler phase speed near periastron in the case of \etacar\ B (see Section \ref{sec:phys}). The only contribution at hard X-rays would then be the emission from secondary particles from hadronic interactions of the freshly accelerated protons.
   
{\bf Maximum particle energies and neutrino fluxes:} 
As noted by \citet{Gupta2017}, neutrino emission from \etacar\ is potentially detectable if high enough particle energies can be reached.
Away from periastron the cutoff energy of the protons is constrained to about
\SI{400}{\giga\electronvolt} by \fermi\ data and HESS limits, see Figure \ref{fig:modelSEDs}. At periastron, the current experimental data only constrain the cutoff energy to be greater than $\approx\SI{500}{\giga\electronvolt}$. In the framework of our model, the cutoff proton energy is limited to $\sim\SI{1}{\tera\electronvolt}$. The spectral energy distribution of neutrino emission from interacting protons peaks at more than an order of magnitude lower energies and would be overwhelmed by the atmospheric neutrino flux~\cite[see e.g.][]{Kappes2007}. 
Even in the extreme case of $\eta_{\rm acc}=\num{1}$ (leading to an exponential cut-off above \SI{12}{\tera\electronvolt} the resulting expected neutrino fluxes, being less than \SI{2e-11}{\erg\per\square\centi\meter\per\second} above \SI{1}{\tera\electronvolt} and less than \SI{3.3e-13}{\erg\per\square\centi\meter\per\second} above \SI{10}{\tera\electronvolt}, make \etacar\ an unlikely source for astrophysical neutrino detection.

    {\bf HESS observations:} The maximum \g-ray energies around periastron are not constrained by observations so far. Existing and future \hess\ detections of \etacar\ will allow to further constrain $E_{\text{max}}$. In our model, strong absorption occurs at periastron (see Figure \ref{fig:modelSEDs}), which is caused by the close vicinity to the stars and the relatively small size of the emission region compared to apastron. A larger size of the ballistic region due to a larger mixing length would reduce the absorption effects and therefore a detection by \hess\ would give information about the location and spatial extent of the emission region.\\
 
The physical conditions in the shocks within the \etacar\ system are
dramatically different from any other well measured particle
accelerating system. As such they represent an extremely promising
laboratory to study the phenomenon of particle acceleration.
The dense and luminous environments provide an interesting comparison with the relatively new class of \g-ray novae \citep{FermiNovae}. This has been previously noted by \citet{VurmMetzger}, who also discuss the connections with WCRs. The strong evidence reported in this paper in support of the hadronic origin of the \fermi\ data for \etacar\ motivates a deeper study into other binary colliding wind systems.

 Of particular interest is the implications for the new measurements on one of the outstanding problems of high-energy astrophysics, namely diffusive shock acceleration and the interplay of the accelerated particle population with its self-excited magnetic fields. The current supernova cosmic-ray origin paradigm for example makes many assumptions regarding the non-linear behaviour of field amplification, and its ultimate role in the energetic particle transport and acceleration. Testing these assumptions with numerical simulations in a self-consistent manner is currently not possible, making \etacar\ a powerful laboratory. While there are many aspects of \etacar's environment that are quite unique, it is nevertheless remarkable that within each complete orbit of the system the shocks sample a broad range of the parameter space relevant to diffusive shock acceleration in galactic supernovae. In this paper, we have shown that magnetic-field geometry, non-linear shock acceleration and magnetic-field amplification, and other plasma-physics considerations are essential to accurately match observations. With future observations and improved statistics, more detailed predictions of the above mentioned processes can be put to the test.

\section{Acknowledgements}
We thank F. Conte for providing the calculations for the cascading of the energy from \g\g-absorption and C. Duffy for cross-checks of \fermi\ analysis. BR gratefully acknowledges valuable discussions with J. Mackey, and also with D. Eichler and D. Burgess during the Multiscale Phenomena in Plasma Astrophysics workshop at the Kavli Institute for Theoretical Physics.

\bibliographystyle{aa}
\bibliography{EtaCar_modelling2}

\appendix

\section{Further \fermi\ Analysis Details}
\label{sec:app-a}

\subsection{Data Selection and Model Parameters}
\label{sec:app-a:par}

The full set of parameters used in the \fermi\ data selection and those used to construct a model of the region are listed in Table~\ref{tab:fermipar}, for an explanation of the parameters please see the Fermi ScienceTools\footnote{\href{http://fermi.gsfc.nasa.gov/ssc}{http://fermi.gsfc.nasa.gov/ssc}} and {\it FermiPy} documentation \citep{Fermipy2017}.

\begin{table}[htb]
    \centering
    \small
    \begin{tabular}{r l}
    \hline \\
    Parameter & Value  \\ \\
    \hline \\
     Data release & P8R3 \\
     IRFs & P8R3\_SOURCE\_V2 \\ 
     ROI Data Width & \SI{10}{\degree}\\ 
     ROI Model Width & \SI{15}{\degree}\\
     Bin Size & \SI{0.1}{\degree} \\ 
     zmax & \SI{90}{\degree}\\ 
     Analysis Coordinate System & GAL\\ 
     Minimum Energy & \SI{80}{\mega\electronvolt}\\ 
     Maximum Energy & \SI{500}{\giga\electronvolt}\\ 
     MET Start & 239557417\\ 
     MET Stop & 583643061\\ 
     MET Excluded (ASASSN-18fv)  & 542144904 - 550885992\\ 
     evclass & 128\\ 
     evtype & 3\\ 
     Galactic diffuse template &  {\it gll\_iem\_v07.fits}\\ 
     Isotropic background component & {\it iso\_P8R3\_SOURCE\_V2\_v1.txt} \\ 
     \fermi\ Catalog & 4FGL ({\it gll\_psc\_v17.fit}) \\ \\
     \hline\\
    \end{tabular}
    \vspace{3mm}
    \caption{The \fermi data selection and model parameters used in this analysis.}
    \label{tab:fermipar}
\end{table}

\subsection{Model Residuals}
\label{sec:app-a:ts}

\begin{figure}[htb]
  \centering 
  \ifdefined\singlecol
     \includegraphics[width=0.8\textwidth,draft=false]{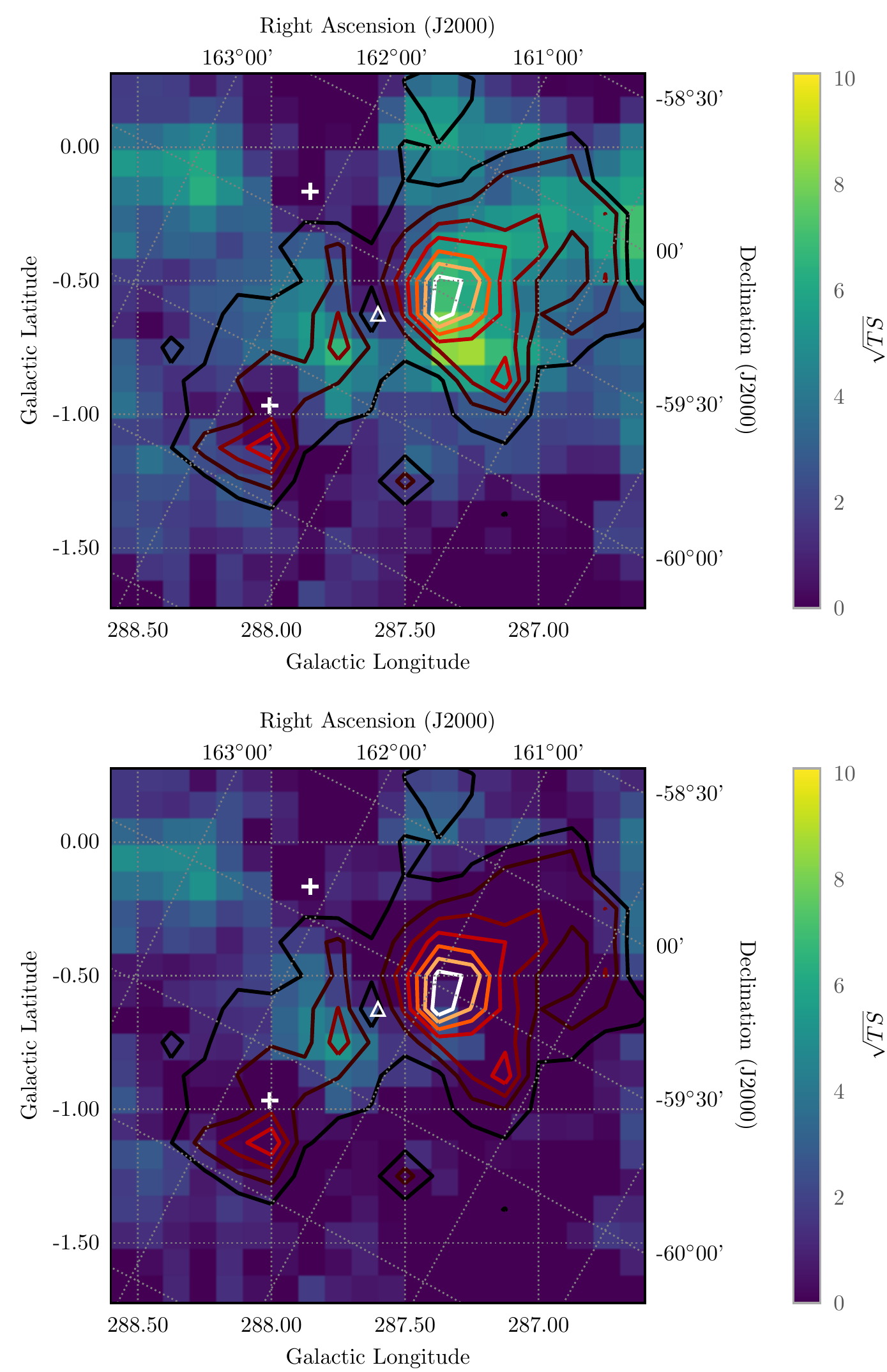} 
  \else
     \includegraphics[width=0.495\textwidth,draft=false]{./plots/tsmap.pdf}
  \fi
  \caption{Residual TS maps produced as a verification step post likelihood fit as part of the analysis chain. The upper map is that resulting from a model with only the default 4FGL catalogue sources included. The position of \etacar\ is marked by a white, hollow, triangle. The two closest 4FGL catalogue sources are indicated by white crosses. Overlaid are contours from a template generated from the CO measurements in \cite{Dame2001}. The lower map shows the residual emission once the CO template is included in the model. 
  }
  \label{fig:tsmap}
\end{figure}

Figure~\ref{fig:tsmap} (upper) shows a residual TS map for a \SI{4}{\degree} by \SI{4}{\degree} region centred on \etacar\ produced as a verification step post-likelihood fit as part of the analysis chain for a model based on only the 4FGL catalog, diffuse Galactic template and isotropic background component. Each point in the map shows the significance that would be attributed to an additional point source at that position. The position of \etacar\ is marked by a white, hollow, triangle. The two closest 4FGL catalogue sources are indicated by white crosses. It is clear that significant residual emission remains when following this standard approach. Overlaid are contours from a template generated from the CO measurements in \cite{Dame2001}, which aligns well with the excess emission. Figure~\ref{fig:tsmap} (lower) shows an equivalent TS map generated following a likelihood fit using a model including this additional diffuse component, indicating the improvement described in Section~\ref{sec:fermi:an}. 

\subsection{Systematic Uncertainties}
\label{sec:app-a:sys}

Systematic uncertainties were estimated for the SED generated from the total \emph{Fermi}-LAT data set by varying analysis parameters and re-running the entire chain. The following parameters were varied:
\begin{itemize}
 \item {\bf Energy range:} Minimum energies between \num{60} and \SI{120}{\mega\electronvolt} and maximum energies of \num{300} and \SI{500}{\giga\electronvolt} were investigated. 
 \item {\bf Region of interest:} The ROI was varied between \SI{5}{\degree}, \SI{10}{\degree}, \SI{15}{\degree} and \SI{20}{\degree}. 
 \item {\bf Starting value for the CO template:} The starting flux of the CO template was varied over an order of magnitude. 
 \item {\bf Free Parameters:} Fits were run with model parameters allowed to vary under a number of conditions. Source normalisation was allowed to vary for sources within \SI{3}{\degree} and \SI{5}{\degree} of \etacar\ with a TS $>\num{10}$, \num{100}, \num{1000}. Shape parameters were also allowed to vary within the same distance of \etacar\ with an additional selection of TS $>\num{100}$, \num{1000}, \num{10000}. 
 \item {\bf SED energy binning:} The SED energy binning was varied between \num{12} and \num{3} bins per decade.
\end{itemize} 

The above variations resulted in hundreds of SEDs. At each point in each SED the fractional error from an interpolated mean SED was determined. The resulting distribution of fractional error as a function of energy was then fitted with a simple cubic polynomial. This function was then used to compute a smooth band around the SED points presented in Figure~\ref{fig:sed} (grey area).

\subsection{Orbital Phase SED Analysis}
\label{sec:app-a:sed}

\begin{figure}[htb]
  \centering 
  \ifdefined\singlecol
     \includegraphics[width=0.8\textwidth,draft=false]{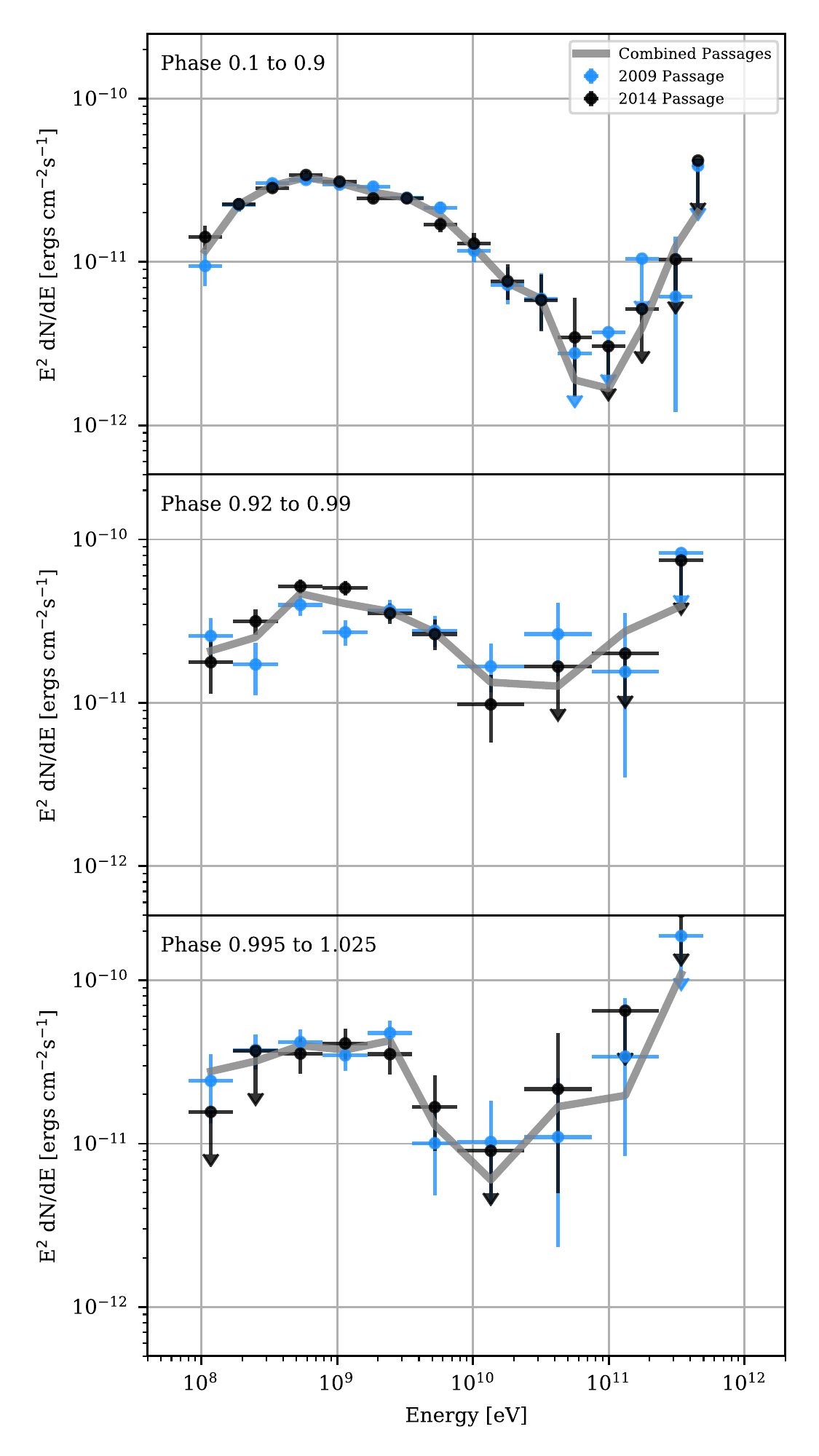} 
  \else
     \includegraphics[width=0.495\textwidth,draft=false]{./plots/perpassage_seds.pdf}
  \fi
  \caption{SEDs from \fermi\ analysis for the same phase ranges presented in Figure~\ref{fig:modelSEDs}. The results from a combined analysis of both orbital periods is given by the solid grey lines. Results from an independent analysis of each orbital passage are shown by the data points (blue: periastron passage \num{2009}, black: periastron passage \num{2014}).
  }   
  \label{fig:passageSEDs}
\end{figure}

In Section~\ref{sec:fermi:phasesed} we describe phase-resolved SED analysis and in Section~\ref{sec:model} the results are presented for a combined analysis of the \fermi\ data from both orbital periods in Figure~\ref{fig:modelSEDs}. Figure~\ref{fig:passageSEDs} shows the sames results (grey solid lines) together with the data points obtained from independently analysing each orbital period. The first orbital period (periastron passage \num{2009}) is shown by the blue data points and the second (periastron passage \num{2014}) by the black data points.

\label{lastpage}

\end{document}